\documentclass[%
 reprint,
showpacs,preprintnumbers,
 amsmath,amssymb,
 aps,
]{revtex4-1}

\usepackage{graphicx}
\usepackage{subfigure}
\usepackage{dcolumn}
\usepackage{bm}
\usepackage{hyperref}
\usepackage{epsfig}
\usepackage{color}

\def\ba{\begin{eqnarray}}
\def\ea{\end{eqnarray}}
\def\bea{\begin{eqnarray}}
\def\eea{\end{eqnarray}}
\def\be{\begin{equation}}
\def\ee{\end{equation}}

\def\({\left(}
\def\){\right)}
\def\[{\left[}
\def\]{\right]}

\begin{document}

\preprint{IGC-18/2-1}
\title{
Gravitational Waves from Binary Mergers of Sub-Solar Mass Dark Black Holes
}

\author{Sarah Shandera$^{1,2}$}
\email{ses47@psu.edu}
\author{Donghui Jeong$^{1,3}$}
\email{djeong@psu.edu}
\author{Henry S. Grasshorn Gebhardt$^{1,3}$}
\email{hsg113@psu.edu}

\affiliation{
$^{1}$Institute for Gravitation and the Cosmos, The Pennsylvania State University, University Park, PA 16802, USA}
\affiliation{
$^{2}$Department of Physics, The Pennsylvania State University, University Park, PA 16802, USA}
\affiliation{
$^{3}$Department of Astronomy and Astrophysics, The Pennsylvania State University, University Park, PA 16802, USA}

\date{\today}

\begin{abstract}
We explore the possible spectrum of binary mergers of sub-solar mass black holes formed out of 
dark matter particles interacting via a dark electromagnetism. We estimate 
the properties of these dark black holes by assuming that their formation
process is parallel to Population-III star formation; except that dark 
molecular cooling can yield smaller opacity limit. We estimate the binary
coalescence rates for the Advanced LIGO and Einstein telescope, and find that
scenarios compatible with all current constraints could produce dark black
holes at rates high enough for detection by Advanced LIGO.
\end{abstract}

\maketitle

{\it Introduction -}\hspace{0.5cm}
Our understanding of dark matter relies almost entirely on observations of how it gravitates on very large scales. The recent gravitational wave detections of binary black hole systems \cite{Abbott:2016blz,Abbott:2016nmj,Abbott:2017vtc,Abbott:2017oio} provide information about gravitating structures on very small scales, and have re-opened a discussion into the possibility that dark matter consists of compact objects which may
be entirely baryonic in origin \cite{Nakamura:1997sm,Bird:2016dcv,Sasaki:2016jop}. Here, we point out that the spectrum of merging compact objects, especially in the sub-solar-mass regime, also constrains a large family of non-baryonic dark matter scenarios. In the event of a detection, the mass of the small black hole would provide a direct constraint on the mass of a dark-sector particle, for example, through the Chandrasekhar limit.

{\it Atomic dark matter -}\hspace{0.5cm} If the cosmologically observed dark matter consists entirely or partly of particles not in the standard model, it may have richer physics than that of the cold dark matter scenario. In particular, dark matter particles that carry one or more new charges may have cooling channels that allow gravitationally bound structures to dissipate kinetic energy into dark radiation. Such scenarios have been studied for many years \cite{Ackerman:mha,Feng:2009mn,Kaplan:2009de,Kaplan:2011yj,Fan:2013yva, Fan:2013tia}, but precise cooling rates for the simplest ``atomic dark matter" scenario have been calculated only very recently \cite{Rosenberg:2017qia, Buckley:2017ttd, DAmico:2017lqj}. Atomic dark matter models are subject to a variety of constraints \cite{Ackerman:mha, CyrRacine:2012fz, Cyr-Racine:2013fsa,Agrawal:2016quu,Ghalsasi:2017jna}, but some viable parameter space remains. 

Here we consider a dark sector consisting of a heavy fermion of mass $m_X$ (a proton analog), a light fermion of mass $m_c$ (a dark electron) and a dark photon. (LIGO constraints are also relevant for dark sectors with more complex particle content \cite{DAmico:2017lqj}.) We denote the dark fine-structure constant as 
$\alpha_D$.
Then, dark matter structures can cool and collapse by processes analogous to those that occur in gravitationally bound clouds of hydrogen. In the absence of dark nuclear physics, the only possible end state for gravitational collapse is a black hole. The minimum mass for a dark-sector black hole (DBH) is set by the Chandrasekhar limit and depends on the mass of the heavy particle as \cite{Chandrasekhar:1931}
\begin{equation}
M^{\rm Dark}_{\rm Chand.}=1.457M_{\odot}\left(\frac{m_p}{m_X}\right)^2,
\end{equation}
where the proton mass is $m_p=0.938$ GeV.  

If dark black holes form, the present-day coalescence rate for black holes with typical mass $M_{\rm DBH}$ in a galactic halo with total dark matter mass $M_{\rm DM}$ can be estimated by
\begin{align}
\label{eq:ndot}
\nonumber
\dot{R}\sim&\left(\frac{M_{\rm DM}\times f_{\rm cool}\times f_{\rm form. \,eff.}}{M_{DBH}}\right)\times f_{\rm binary}\\
&\times\left.\left[\frac{dP(T_{\rm merge})}{dT_{\rm merge}}\right]\right|_{T_{\rm merge}\sim 10^{10}{\rm yr}}
\end{align}
where $f_{\rm cool}$ is the fraction of dark matter that can dynamically cool, $f_{\rm form. \,eff.}$ is the fraction of the cooling dark matter that ends up in DBHs, $f_{\rm binary}$ is the number of binary systems 
compared to total DBHs, and $dP(T_{\rm merge})/dT_{\rm merge}$ is the probability density of the merger times of the binaries.

In the rest of the paper, we demonstrate that there is no obvious obstruction to the formation of sub-solar-mass DBHs. Since DBH formation is likely to share many features of Population-III (Pop III) star formation, where no metal is involved and there are fewer complex feedback mechanisms generated by nuclear physics, we use the literature on formation and binary parameters of Pop III stars to estimate each of the terms in Eq.(\ref{eq:ndot}). For several choices for dark
sector parameters that are broadly consistent with current constraints, we estimate the event rates for Advanced LIGO operating at current and designed
sensitivity as well as for a Einstein-telescope like future gravitational wave observatory.

{\it Cooling, fragmentation, and collapse in the dark sector - }
Many of the relevant atomic cooling processes for dark matter charged under a dark electromagnetism have been recently calculated in detail by Rosenberg and Fan \cite{Rosenberg:2017qia}. As demonstrated by Buckley and DiFranzo \cite{Buckley:2017ttd}, a choice of parameters for the dark U(1) sector (as well as $\xi=T_D/T_{\rm CMB}$, the ratio of the temperature of the dark sector to that of the visible photon--denoted as CMB) defines the range of halo masses which can cool by
processes analogous to those for hydrogen gas. These processes require a certain minimum density to be operative, so they need not alter the formation of usual cold dark matter halos on very large ($\simeq {\rm Mpc}$) scales.

At the same time, the coupling of dark matter to dark radiation suppresses
structure on scales smaller than the sound horizon scale at the kinematic
decoupling time of dark species. This dark acoustic oscillation (DAO) scale provides an approximate lower limit on the size of substructures that can 
form. In order for all of dark matter to be charged under the dark force, with masses and couplings such that DBHs likely to be accessible with Advanced LIGO are formed, while still remaining consistent with DAO constraints \cite{Cyr-Racine:2013fsa} we assume $\xi=T_D/T_{\rm CMB}=0.02$. Larger $\xi$ will further suppress small-scale structure and lead to a smaller fraction of dark matter that can collapse, and so a smaller number of DBHs. Assuming only the degrees of freedom in the standard model and that the dark and baryonic sectors were in thermal equilibrium gives $\xi=0.5$, so some additional physics (e.g. differential reheating \cite{Adshead:2016xxj}) is required to achieve this number. 

We follow \cite{Rosenberg:2017qia, Buckley:2017ttd} to estimate the range of halo masses for which the time for the halo to lose order one of its energy by cooling is shorter than the free-fall time scale, including inverse-Compton scattering of CMB photons, free-free scattering of dark electrons, free-bound scattering, and collisional excitation of dark hydrogen (1s$\rightarrow$ 2p). 
In order not to spoil the large-scale structure formation, we choose example parameter sets by setting the maximum halo size that can cool significantly in a free-fall time to $10^{11}$ M$_{\odot}$. From the study of sub-structure in the Via Lactea simulation \cite{Madau:2008fr}, we estimate that these parameters would place no more than a few percent of the dark matter in the Milky Way in the cooling regime. More precise estimates would require a suite of
dedicated numerical simulations, so we take $f_{\rm cool}\sim 0.01$ as a typical cooling fraction. 

For halos in the right mass range to support cooling, the minimum Jeans mass can be estimated using the opacity-limit argument, which gives \cite{Rees:1976, Low:1976}
\begin{equation}
\label{eq:mjmin}
M_{J,min}\propto \left(\frac{m_p}{m_X}\right)^{9/4}\left(\frac{T}{10^3 K}\right)^{1/4}M_{\odot},
\end{equation}
where $m_p$ is the proton mass. The minimum Jeans mass decreases faster with increasing $m_X$ than the Chandrasekhar limit does. 
The Pop III stars form in small halos with virial temperatures around $10^3K$, 
which is cooled to around 200 $K$ by the rotational line cooling 
of molecular hydrogen \cite{Abel:2001pr,Bromm:2001bi}.%

For dark molecular hydrogen, the separation between the ground state and the first excited state scales as
 \begin{equation}
 \label{eq:DMH}
 \Delta E=\left(\frac{m_p}{m_X}\right)\left(\frac{m_c}{511\,{\rm keV}}\right)^2\left(\frac{\alpha_D}{0.0073}\right)^2\times 512\,K.
 \end{equation}  
 Numerical simulations of Pop III stars find that the collapsing gas creates $\sim 1M_{\odot}$ protostars which accrete rapidly to form relatively massive stars with birth masses of order $100 M_{\odot}$ \cite{Abel:2001pr, Bromm:2003vv}.
We use Eq.(\ref{eq:DMH}) for the temperature in Eq.(\ref{eq:mjmin}), fixing the constant of proportionality from Pop III studies. Then, assuming the dynamics of DBH formation is not too different from the formation of Pop III protostars, we estimate that the minimum mass of DBHs at formation will be around
\begin{equation}
M_{\rm DBH,min}\sim \left(\frac{m_p}{m_X}\right)^{9/4}\left(\frac{T}{10^3 K}\right)^{1/4} 10^3 M_{\odot}.
\end{equation}
This estimate does not account for the fact that both baryonic and dark 
matter will be present in the cloud, and it assumes there is no 
coupling between the two sectors other than gravity. 

Once these small black holes form, they are likely to stay small. The Eddington accretion rate scales as $\dot{M}=\frac{4\pi G Mm_X}{\epsilon c}\left(\frac{3}{8\pi}\right)\left(\frac{m_c c^2}{ \alpha_D\hbar c}\right)^2$, where $\epsilon$ is the fraction of potential energy from infalling matter that can be radiated away as heat. 
In the cases studied here, there is likely to be no appreciable accretion from dark sector or baryonic matter after the formation period.

{\it Modeling the dark black hole population and coalescence rate -}\hspace{0.5cm} 
\begin{figure}[t!]
\centering
\includegraphics[width=.51\textwidth]{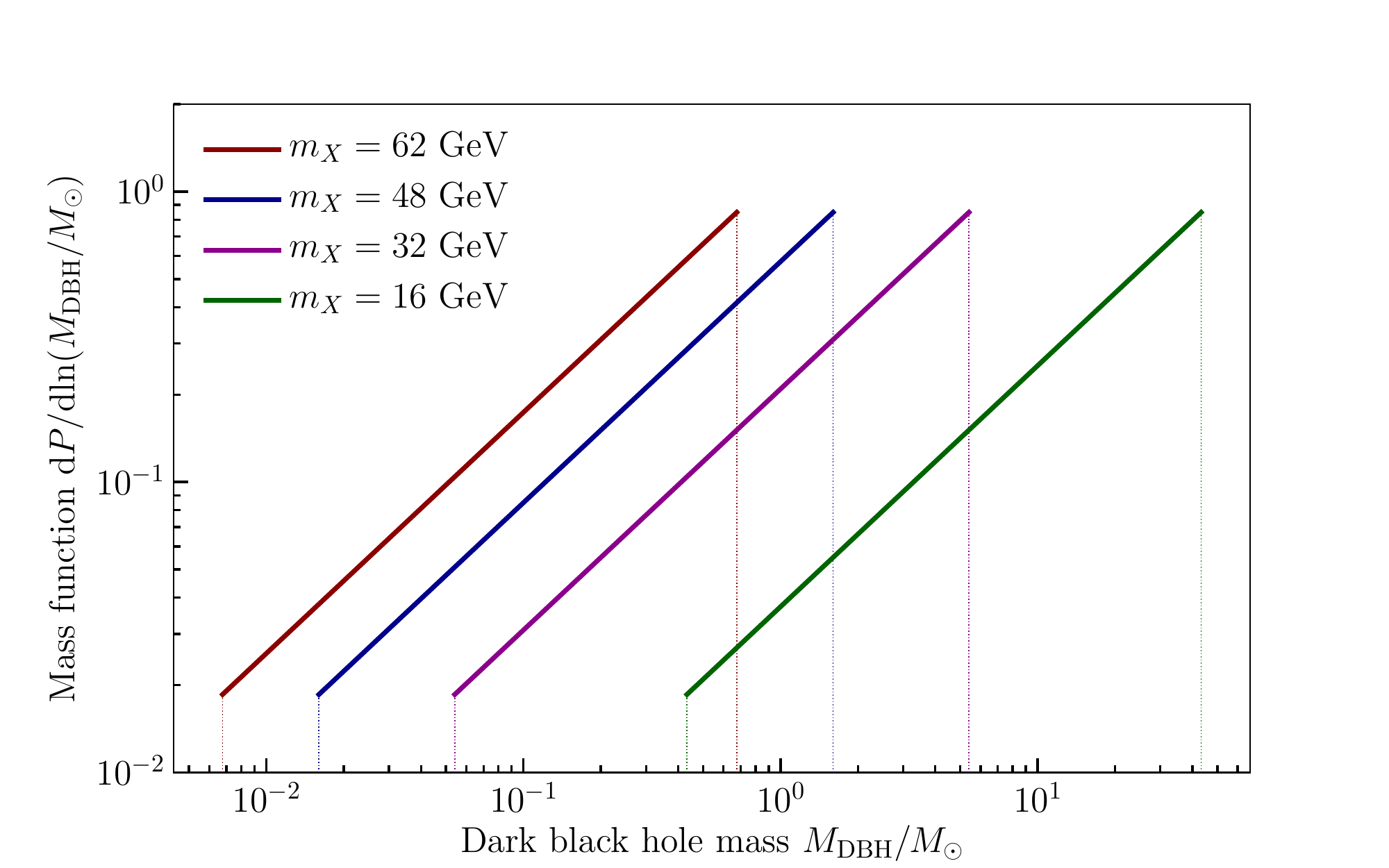}
\caption{\label{fig:whatcools} 
The mass function of black holes for four dark matter masses that 
we consider here. For all cases, we assume 
$\xi\equiv T_D/T_{\rm CMB}= 0.02$, $\alpha_D=0.01$, and set $m_c$ so that 
the dissipation does not affect the formation of dark matter halos above 
$10^{11}\,{M_\odot}$.
\label{fig:nM}
}
\end{figure}
%
In simulations of Pop III star formation, about 0.1\% of the gas ends up in stars \cite{Stacy:2012iz}. Conservatively, we assume an efficiency factor of $f_{\rm form. \,eff.}=10^{-3}$, to relate the fraction of dark matter that is in collapsing structures to the fraction that actually ends up in black holes. Together with the $f_{\rm cool}=0.01$, this would put $10^{-5}$ of the dark matter in dark black holes. However, simulations and observations of the stellar mass fraction today find that stars make up a few percent of the total halo mass (see, eg, \cite{Pillepich:2017fcc} and references therein), which means that more than 10\% of baryons in the halo are in stars. The combination $f_{\rm cool}\times f_{\rm form. \,eff.}$ may be plausibly as high as $10^{-3}$. We take this as the optimistic case.

\begin{figure*}[th]
\centering
\includegraphics[width=.49\textwidth]{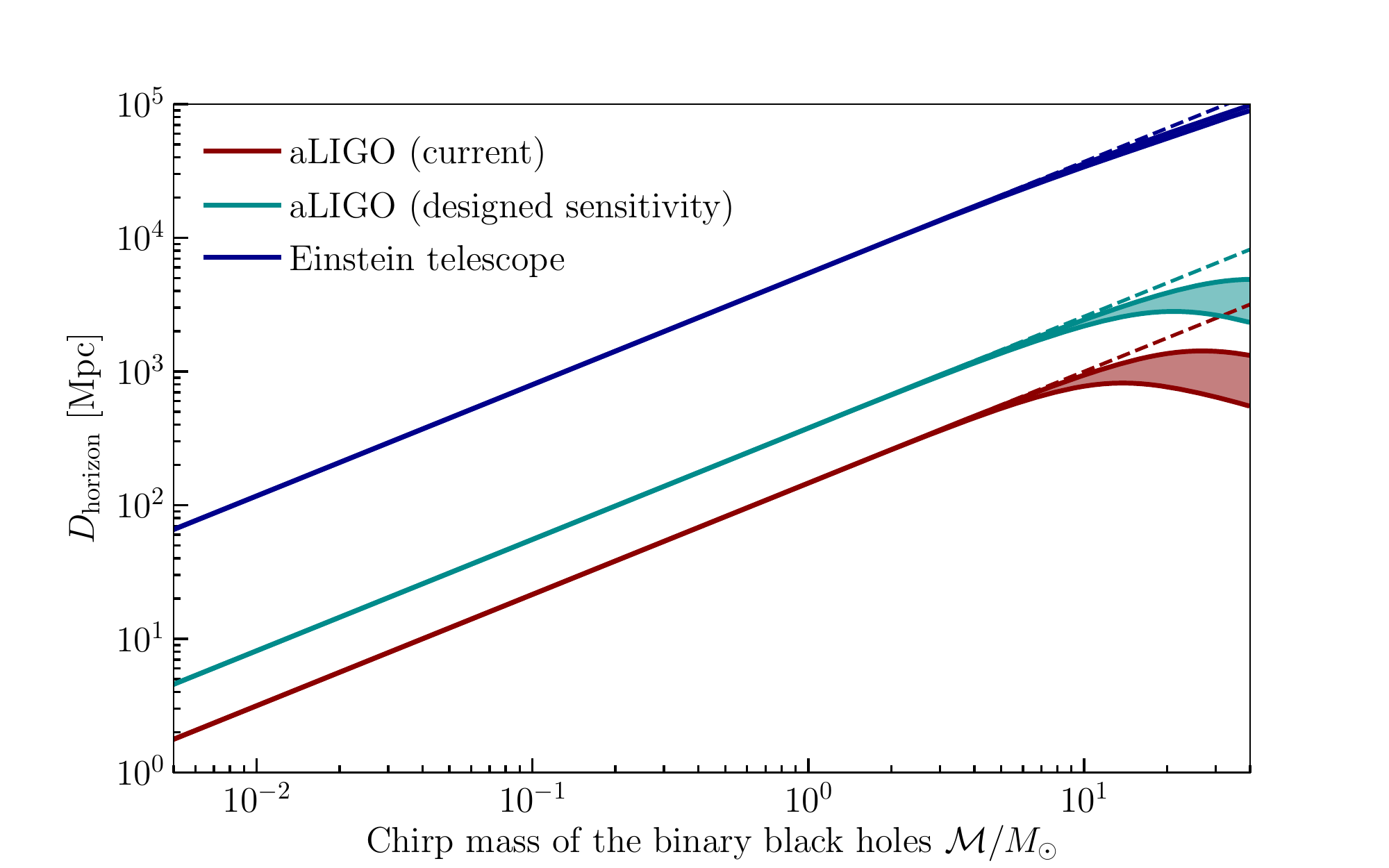}
\includegraphics[width=.49\textwidth]{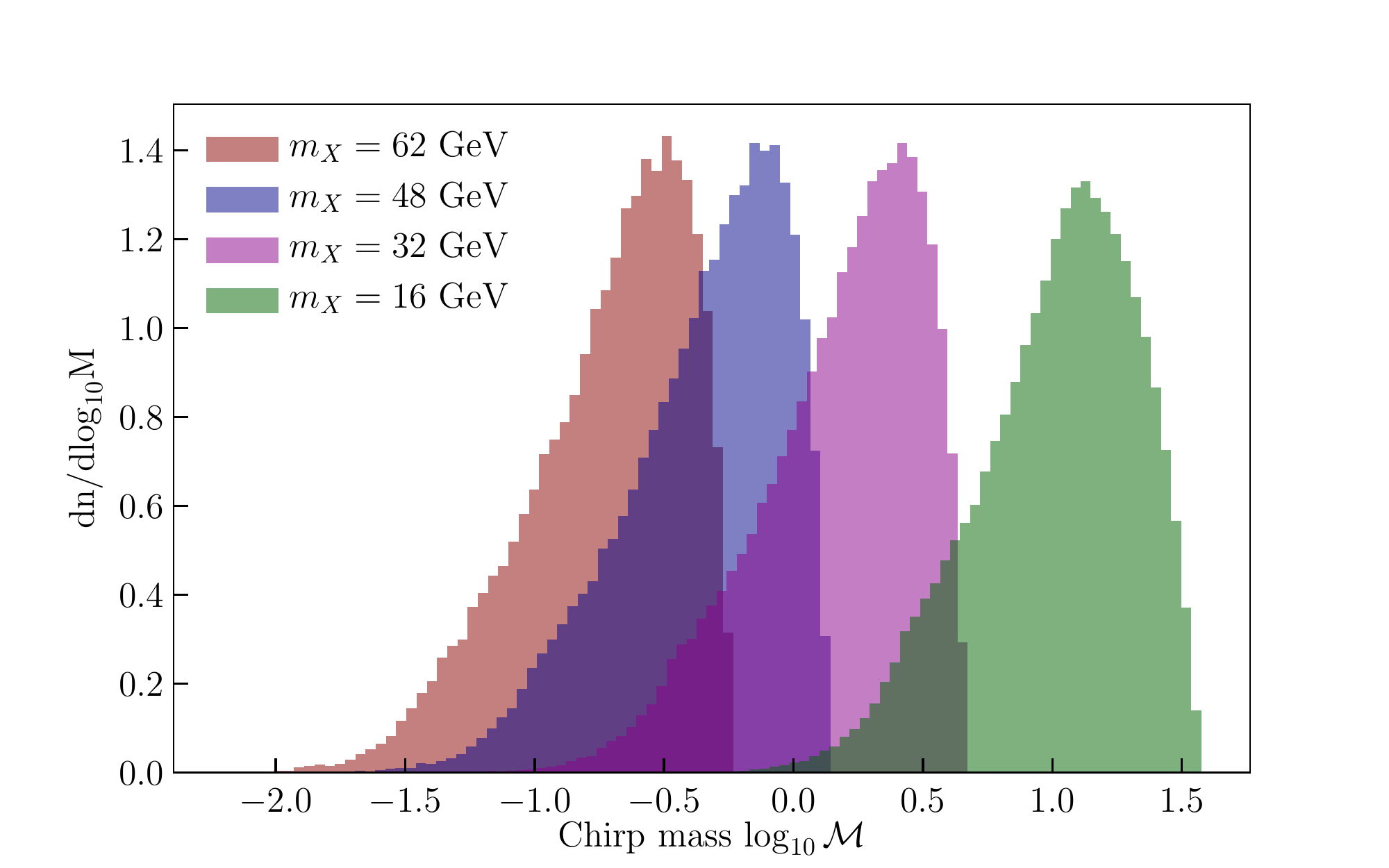}
\caption{\label{fig:distance}
({\it Left:})
The horizon distance (maximum luminosity distance) for the 
8-$\sigma$ detections from the current advanced LIGO (aLIGO), 
advanced LIGO at design sensitivity, and Einstein telescope,
as a function of chirp mass ${\cal M}\equiv [q^3/(1+q)]^{1/5}m$. For each case, upper (lower) solid line is for 
$q=1$ ($q=0.1$), and dashed line is the approximation 
$D_{\rm horizon}({\cal M}) \propto 
({\cal M}/M_\odot)^{6/5}$ with proportionality
constant 147, 378, 5450 {\rm Mpc} for the current aLIGO, aLIGO at 
design sensitivity, and Einstein telescope, respectively.
({\it Right:})
The distribution of the chirp mass ${\cal M}$ of the DBH binaries merging today, for the 
four dark matter parameter choices shown in Table \ref{table}.
}
\end{figure*}
%
We use the mass function and binary parameters from observational and numerical studies of Pop III stars as a guideline for the DBH population. We assume a mass function of the form $dP_m\propto m^{-b}$, with birth masses that range over two orders of magnitude. Some studies suggest that the initial mass function of Pop III stars is significantly flatter than the Salpeter form ($b=2.35$). The range of possible values includes $b=-0.17$ as fit to the simulations results
of \cite{Greif:2011} in \cite{Stacy:2012iz}, or log-flat ($b=1$), as used \cite{Hartwig:2016nde}. We use $b=0.17$, the best fit from \cite{Stacy:2012iz}, for our numerical results. Figure \ref{fig:nM} shows the DBH mass function for several choices of the dark proton mass, $m_X$.

We use the binary parameters reported by \cite{Stacy:2012iz}: 
fraction of stars in binaries, $f_{\rm binary}=0.26$ 
(corresponding to their $f_B=0.36$),
and the mass ratio ($q=m_{\rm light}/m_{\rm heavy}$) distribution 
$P_q\propto q^{n_q}dq$ with $n_q=-0.55$. 
We assume a thermal distribution for the eccentricity \cite{Jeans:1919}, $e$, so that $dP_e\propto e^{n_e} de$ with $n_e=1$. We take $0.1<e<1$ \cite{Hartwig:2016nde}. For the distribution for the semi-major axis, $a$, we follow \cite{Hartwig:2016nde} and use $dP_a\propto x^{n_a} dx$, with $x={\rm Log}_{10}(a/a_*)$, $n_a=-1/2$ and $a_*$ chosen to preserve the shape of the distribution. Hartwig et al \cite{Hartwig:2016nde} consider $0.23 AU<a<9300AU$ for objects between 3
$M_{\odot}$ and 300 $M_{\odot}$, with $a_*$=R$_{\odot}\approx 0.0047$ AU. Assuming that the separation between objects scales approximately as $M^{1/3}$, we take, for example, $0.06 <a<\, 2400$ AU as a most probable range 
when the minimum mass of DBH is 0.054\,$M_\odot$ ($m_X=100$ GeV case).

The time for a binary system of objects with masses $m_1$ and $m_2$ to merge due to loss of energy to gravitational radiation is given approximately by the Peters formula \cite{Peters:1963,Peters:1964}. For eccentricity $e$, and semi-major axis $a$
\begin{equation}
T_{\rm merge}=  \frac{(3\times10^{9}{\rm yr})M_{\odot}^3}{m_1m_2(m_1+m_2)}\left(\frac{a}{0.01\;{\rm AU}}\right)^4(1-e^2)^{7/2}.
\end{equation}
Using the distributions for the binary parameters above, together with the Peters formula, we estimate the merger rate today, for a Milky Way 
equivalent galaxy (MWEG) defined by total mass of 
$10^{12}M_{\odot}$ \cite{eadie/etal:2017} assuming the global dark 
matter fraction of 84\%.

We then compute the detection rate, $\dot{N}$, for the gravitational wave
observatories from the yearly merger rate per MWEG (called {\it raw} rate
in Table \ref{table})
by \cite{Abadie:2010cf}
\begin{equation}
    \dot {N}=\left(\frac{\#{\rm mergers}}{{\rm MWEG\; yr}}\right)\frac{4}{3}\pi\left(\frac{D_{\rm horizon}}{{\rm Mpc}}\right)^3(2.26)^{-3}(0.0116)
\end{equation}
where the last two numerical factors account for inhomogeneous coverage of sky position and orientations, and the expected density of MWEGs in the coverage area.

The horizon distance $D_{\rm horizon}$ is typically calculated by finding the distance at which a signal-to-noise ratio of 8 is achieved. 
We compute the horizon distance by using the inspiral portion of the merger signal (scaling with frequency as $f^{-7/6}$ \cite{Abadie:2010cf}) 
and the noise spectral density curves for current and design 
sensitivity of Advanced LIGO, as well as the Einstein telescope. 
The horizon distance as a function of chirp mass of the binary 
${\cal M}^5\equiv \mu^3M^2$ ($\mu$ is the reduced mass and $M$ is the 
total mass) is shown in the left panel of Figure \ref{fig:distance}. 
When computing the event rate for the Advanced LIGO and Einstein telescope,
we average over the cube of the horizon distance for each binary in our population with a merger time between 9 and 10 Giga-years. 
The right panel of Figure \ref{fig:distance} shows the distribution of chirp masses for present-day mergers. 
We present the event rates in Table \ref{table}. 
Table \ref{table} shows the conservative (optimistic) rate estimates for several dark matter scenarios that could result in DBHs with masses accessible by 
Advanced LIGO and Einstein telescope. Conservative (optimistic) rates assume $f_{\rm cool}\times f_{\rm form. \,eff.}=10^{-5}$ ($10^{-3}$).
Note that the actual rate of interest will be higher if we include the possibility of binaries containing one sub-solar DBH and one standard black hole.

\begin{center}
    \begin{table*}
        \begin{tabular} {|c|c|c|c|c|c|c|c|c|c|}
\hline
$m_X$ & $m_c$ & $M^{\rm dark}_{\rm Chand.}$ &$M_{\rm DBH}$& \multicolumn{4}{c|}{Rates per year} & $m_1<1.4$ & $m_1,m_2<1.4$ \\
\cline{5-8}
[GeV] & [keV] &  $[10^{-5}M_\odot]$ & $[M_\odot]$ 
& raw (MWEG$^{-1}$)
& aLIGO (current) 
& aLIGO (full)
& Einstein T.
& [\%]  & [\%] \\
\hline
 62 & $31$ &  $33$ & $0.0068- 0.68$  & 2.0$\times 10^{-6} (10^{-4})$ & 
 0.0012 (0.12) &
 0.020 (2.0)   &
 60 (6000)     &
 100\% & 100\%  \\ 
\hline
48 & $47$ &  $56$ & $0.016-1.6$  & 1.3$\times 10^{-6} (10^{-4})$ & 
 0.0065 (0.65) &
 0.11   (11)   &
 330   (33k) &
 99\%   & 79\%          \\ 
 \hline
 32& $70$ &  $125$ & $0.054 - 5.4$  &$6.6\times 10^{-7} (10^{-5})$& 
 0.068 (6.8)  &
 1.1   (110)  &
 3500  (350k)&
 53\% & 9.3\%\\ 
\hline
 16 & $140$ &  $500$ & $0.43-43$  &$1.9\times 10^{-7} (10^{-5})$ &
 0.89  (89)   &
 22    (2200) &
 92k (9200k) &
 9.8\% &0.14\%\\ 
\hline
\end{tabular}
\caption{
    DBH masses and binary merger rates today, estimated using the procedure in the text, for several choices of dark proton mass $m_X$ and dark electron mass $m_c$. All black hole masses are given in solar masses. In all cases we have set the dark fine structure constant to $\alpha_D=0.01$ and the ratio of present day temperature of the dark sector to photon temperature to $\xi=0.02$. The conservative (optimistic) rates use $f_{\rm cool}\times f_{\rm form. \,eff.}=10^{-5} (10^{-3})$.
    The optimistic rate for $m_X=16$ GeV is high enough it may already be constrained by existing LIGO data. The last two columns show the percent of binaries where one or both black holes in the binary has a mass less than the standard Chandrasekhar mass (1.4 $M_\odot$).
}\label{table}
\end{table*}
\end{center}

{\it Discussion - } Our estimates show that reasonable parameter regimes for the dark sector may give populations of obviously non-baryonic black holes (that is, with masses below the baryonic Chandrasekhar limit) within 
the reach of future instruments (e.g., the Einstein Telescope \cite{Hild:2010id}) and possibly even Advanced LIGO. By assuming the most favorable possible merger rate, a null result for sub-solar mass black hole searches from gravitational waves can provide relatively direct constraints on the fraction of dark matter in sub-solar mass black holes. Through dedicated numerical studies to refine the estimates presented here, gravitational wave data can constrain particle physics parameters of an ``atomic" dark sector.

LIGO did carry out a search in the range of $0.2-1.0 M_{\odot}$ in 2005, constraining the population to be fewer than 63 per MWEG per year (90\% confidence). Historically, the only motivation for such a search has been a possible primordial black hole population, which may be distinguishable from the DBH scenario by LIGO itself if sufficient spin information can be acquired.

Note that even in our optimistic case, DBHs make up just $0.1\%$ of dark matter and have an extended mass function, so the population evades existing micro-lensing constraints \cite{Alcock:2000ph, Tisserand:2006zx,Zumalacarregui:2017qqd} which come in when compact objects make up order 10\% of dark matter. If future microlensing searches similar to \cite{Kains:2017} target small black holes, that may be a promising way to detect or constrain the population. Although these black holes are small, if they are sufficiently clustered X-ray emission from their accretion disks may also be detectable. 

While the DBH population is presently not well constrained, interacting dark matter scenarios also alter small-scale structure. Studies of nearby galaxies find good agreement with the predictions from cold dark matter simulations \cite{Natarajan:2017sbo}, but are possibly not sensitive to shifting about $\sim 1\%$ of the dark matter into the cooling regime. Simulations of atomic dark matter in the literature include work claiming rather tight constraints from sub-structure, albeit in a slightly different parameter space than we have used here, from various features of substructure \cite{Ghalsasi:2017jna} and work showing that atomic dark matter may alleviate small-scale structure tensions  \cite{Boddy:2016bbu}. If the relationship between $m_X$ and the minimum DBH mass used here survives a more careful analysis, a sub-solar mass black hole search with LIGO would provide complementary constraints on the region of parameter space that is on the edge of being excluded by current self-scattering bounds in the literature \cite{Agrawal:2016quu}. Even if these constraints ultimately imply that only a fraction of dark matter can be atomic, scenarios where all the atomic dark matter can cool may still lead to a DBH population of comparable size and statistics to the one we have considered here. 

{\it Acknowledgments - } We thank Chad Hanna for inspiring us to consider novel black hole populations, Anne-Sylvie Deutsch for collaboration on early stages of this project, Mike Eracleous for discussions of astrophysical constraints, and B.S. Sathyaprakash for supplying noise spectral density estimates for aLIGO and the Einstein Telescope. We thank M. Buckley for communication regarding his paper. We thank Yu-Dai Tsai for providing several useful comments on version one of this paper, and for spotting a typo in the $m_X$ values quoted in the table.
\bibliographystyle{unsrt}

\end{document}